# Application of Differential Equations in Projecting Growth Trajectories

### Ron W. Nielsen[1]

## Abstract

**Mathematical method based on a direct or indirect analysis of growth rates is described. It is shown how simple assumptions and a relatively easy analysis can be used to describe mathematically complicated trends and to predict growth. Only rudimentary knowledge of calculus is required. Projected trajectories based on such simple initial assumptions are easier to accept and to understand than alternative complicated projections based on more complicated assumptions and on more intricate computational procedures. Examples of the growth of population and of the growth of the Gross Domestic Product are used to illustrate the application of this method of forecasting.**

## 1. Introduction

Mathematical analysis of trends could be either complicated or oversimplified. Complicated analysis could be based on Monte Carlo simulations (Rubinstein & Croese, 2016) or on statistical modelling (Hyndman, & Arthanasopoulos, 2014; Makidakis, Wheelwright & Hyndman, 1998) while oversimplified analysis is often based on fitting a straight line or some other distribution *directly* to the time-dependent series. Mathematical distributions fitted directly to data could be used for the *description* of data but they have limited application in forecasting. In particular, higher-order polynomials are inapplicable because their parameters depend strongly on the range of the time series. Complicated analysis based on Monte Carlo calculations or on statistical modeling could be considered more reliable but by being more complicated they are less accessible and perhaps even less appealing.

I want to suggest an alternative method of analysis of time-dependent series and of forecasting, the method based on the mathematical analysis of *growth rates*. The proposed method is supported by the fundamental observation that fluctuations and oscillation in the growth rates have no essential impact on the associated trends (Nielsen, 2016a, 2016b). The analysis of growth rates can be significantly simplified. They, or their suitably defined substitutions, can be often fitted even by a straight line. By using rudimental differential calculus, such simple initial description of data can be then translated into mathematically more complicated descriptions of trends and used in forecasting.

[1] AKA Jan Nurzynski, Environmental Futures Research Institute, Griffith University, Gold Coast Campus, Qld, 4222, Australia.
Email: r.nielsen@griffith.edu.au, ronwnielsen@gmail.com

In science, descriptions and explanations based on simple assumptions are always preferable and more acceptable then explanations based on complicated assumptions and on complicated mathematical procedures. Trends could be complicated but it does not mean that their description has to be based on complicated assumptions. It is easy to introduce a series of complicated assumptions and then to derive complicated descriptions of trends. The real challenge is to use the simplest possible assumptions and yet to describe even the most complicated trends. The aim of this paper is to show how to do it.

It is always advisable to reduce the analysis of data to a straight line, for three reasons: (1) the straight line is the simplest mathematical function, (2) any deviation of data from the straight line can be easily detected, and (3) the parameters describing a straight line do not depend critically on the range of data. Straight-line description of data includes also exponential representation because exponential function can be easily reduced to a linear function by using a logarithm of analysed data.

I am going to show how the analysis of trends can be reduced to such a simple representation of time-dependent series and how the parameters based on such a simple assumption can be used to describe even mathematically more complicated trend trajectories.

## 2. Mathematical method – fundamentals

Growth rate is defined by the following equation:

$$R \equiv \frac{1}{S}\frac{dS}{dt} \tag{1}$$

where $S(t)$ is the size of the growing entity and $t$ is the time.

For the direct calculations from data it is:

$$R_{i+1} = \frac{1}{S_i}\frac{S_{i+1} - S_i}{t_{i+1} - t_i}. \tag{2}$$

The size $S$ can represent, for instance, the Gross Domestic Product (GDP) or the size of the population.

Application of the analysis of growth rates in predicting growth should be obvious, because growth rate determines what we can expect in the future. For instance, exponential growth is characterized by a constant growth rate. Such a growth is unsustainable over a sufficiently long time. Consequently, if we can see that the growth rate describing, for instance, economic growth varies around a constant value, we can take such a pattern as a warning sign, because such a growth will inevitably have to slowed down.

It is obvious, therefore, that if the growth rate is not constant but increases with time, then such a growth is even worse, because it will become unsustainable even faster.

The only growth, which can be sustainable over a long time, or even indefinitely when properly regulated, is the growth characterized by a decreasing growth rate. Such a growth will reach a maximum if the growth rate is going to decrease to zero or it will reach a certain equilibrium level if the growth rates is going to approach asymptotically it zero value.

Thus, even without carrying out any elaborate calculations we can predict whether the growth is likely to be sustainable or not. However, mathematical analysis of growth rates can help in more accurate projections of growth. We can then tell not only whether growth is likely to be sustainable or unsustainable but also to study more closely the predicted trajectories. Such studies can help in determining how soon projected trajectory might become unsustainable or how to regulate the growth rate to reach a desired maximum or the desired level of equilibrium and when.

Mathematical method, described here, has a general application and can be used for any type of data, as long as they are over a sufficiently large range of independent variable to allow for the calculations of the growth rate by using the eqn (2). The described method can be used for any type of growth. If we have a sufficiently large number of data, we can use them to determine the empirical growth rate, analyse it mathematically, use the differential calculus to translate results of such analysis into the description of trends and use the mathematically determined distributions in forecasting.

If analysis of data is carried out by using an appropriately-defined distribution $F(S)$, rather than $S$, then the starting point is to calculate the growth rate of $F(S)$:

$$R \equiv \frac{1}{F(S)} \frac{dF(S)}{dt} \approx \frac{1}{F(S)} \frac{\Delta F(S)}{\Delta t}. \qquad (3)$$

Again, we can use eqn (2) to calculate the growth rate $R$ but now we shall be using $F(S)$ rather than $S$ in this equation. Whether we are using $S$ or $F(S)$ in such calculations, they will usually produce strong fluctuations of $S$. If the data are of exceptionally good quality, fluctuations will be small or negligible. However, in general they will be significantly large and they will be obscuring the general trend of the growth rate. We can still fit a straight line to such fluctuating growth rate and use it to predict growth, but if we want to unravel the general trend of the growth rate we would have to eliminated the obscuring effects of local gradients $\Delta S / \Delta t$ or $\Delta F / \Delta t$ by polynomial interpolation.

Growth rate can be presented as a function of time or as a function of the size of growing entity. We shall now discuss these two possibilities. Fundamental application of differential equations in the description of trends and in forecasting is summarized in Table 1.

If the empirically-determined growth rate can be described by a certain *time*-dependent function $f(t)$, i.e. if

$$\frac{1}{S} \frac{dS}{dt} = f(t), \qquad (4)$$

then to find the mathematical representation of data we have to solve the following differential equation:

$$\frac{dS}{S} = f(t)dt. \qquad (5)$$

Its solution is

$$S(t) = \exp\left[\int f(t)dt\right]. \qquad (6)$$

If

$$f(t) = r = const, \qquad (7)$$

the solution of the eqn (4) is given by the exponential function,

$$S(t) = Ce^{rt}, \qquad (8)$$

where $C$ is related to the constant of the integration. The eqn (8) describes exponential increase, if $r > 0$ or decrease if $r < 0$.

If the empirically-determined growth rate can be represented by a straight line, i.e. if

$$f(t) = a + bt, \qquad (9)$$

where $a$ and $b$ are constants, then

$$S(t) = C \exp\left[ at + 0.5bt^2 \right].\tag{10}$$

In this case, the gradient of $S(t)$ is

$$\frac{dS(t)}{dt} = C(a + bt) \exp\left[ at + 0.5bt^2 \right].\tag{11}$$

For a suitable combination of parameters $a$ and $b$, the distribution given by the eqn (10) will reach a maximum when $a + bt = 0$, i.e. at $t = -a/b$.

If the empirically-determined growth rate can be described by a certain *size*-dependent function $f(S)$, i.e. if

$$\frac{1}{S}\frac{dS}{dt} = f(S),\tag{12}$$

we can express this equation as

$$\frac{dS}{S \cdot f(S)} = dt.\tag{13}$$

We now have a mathematically more complicated problem, because there is no single prescription for the solution of such differential equations.

In the simplest case when $f(S) = r = const$ the solution is again represented by an exponential function. If we take the next least complicated step and assume that $f(S)$ is represented by a straight line, i.e. if

$$f(S) = a + bS,\tag{14}$$

then we have the following differential equation:

$$\frac{dS}{S(a + bS)} = dt.\tag{15}$$

To find how to integrate the left-hand side of this equation let us consider a general case:

$$\frac{dx}{(a + bx)(c + ex)}\quad,\tag{16}$$

where $a$, $b$, $c$ and $e$ are constants.

To integrate this fraction, we split it into two fractions:

$$\frac{1}{(a + bx)(c + ex)} = \frac{A}{(a + bx)} + \frac{B}{(c + ex)},\tag{17}$$

where $A$ and $B$ are certain constants, which we now have to determine.

The right-hand side of the eqn (17) can be expressed as

$$\frac{A}{(a + bx)} + \frac{B}{(c + ex)} = \frac{(c + ex)A + (a + bx)B}{(a + bx)(c + ex)}.\tag{18}$$

By comparing eqns (17) and (18) we can see that

$$(c + ex)A + (a + bx)B = 1,$$ (19)

which gives us a set of two equations:

$$cA + aB = 1,$$ (20a)
$$eA + bB = 0.$$ (20b)

Their solution is

$$A = \frac{b}{\Delta},$$ (21a)

$$B = -\frac{e}{\Delta},$$ (21b)

where

$$\Delta = cb - ae.$$ (22)

So now, the eqn (17) can be replaced by

$$\frac{1}{(a + bx)(c + ex)} = \frac{b}{\Delta}\frac{1}{(a + bx)} - \frac{e}{\Delta}\frac{1}{(c + ex)}.$$ (23)

Integration of the left-hand side of this equation is replaced by the integration of two simpler fractions. Their integration can be done by substitutions. Thus, for instance if we use $u = a + bx$ we get

$$\int \frac{1}{a + bx} dx = \frac{1}{b}\int \frac{du}{u} = \frac{1}{b}\ln u = \frac{1}{b}\ln(a + bx).$$ (24)

Consequently,

$$\int \frac{dx}{(a + bx)(c + ex)} = \frac{1}{\Delta}\ln \frac{a + bx}{c + ex}.$$ (25)

We have derived a useful general formula of integration. In particular, we can see now that

$$\int \frac{dx}{x(a + bx)} = -\frac{1}{a}\ln \frac{a + bx}{x},$$ (26)

because $c = 0$, $e = 1$ and consequently $\Delta = -a$.

We are now ready to solve the eqn (15). Integration of both sides of the equation

$$\int \frac{dS}{S(a + bS)} = \int dt$$ (27)

gives

$$-\frac{1}{a}\ln\frac{a+bS}{S} = t + C,$$   (28)

where $C$ is the constant of integration.

Simple arithmetical manipulations lead to the following solution of the eqn (15):

$$S = \left[ Ce^{-at} - \frac{b}{a} \right]^{-1}.$$   (29)

Constant $C$ can be determined by normalising calculated $S$ to data at a certain time $t_0$,

$$C = \left[ \frac{1}{S_0} + \frac{b}{a} \right] e^{at_0},$$   (30)

where $S_0$ is the empirical size of the growing entity (e.g. the GDP) at a selected time $t_0$.

If $a + bS = r = const$, i.e. if the growth rate is constant, the eqn (29) gives exponential growth.

If $a + bS \neq const$ we have two possibilities: the growth rate represented by $a + bS$ can either increase or decrease with the size of the growing entity:

$$\frac{1}{S}\frac{dS}{dt} = a + bS.$$   (31)

If $b < 0$, the eqn (29) represents the logistic-type of growth. The characteristic signature of this type of growth is its linearly decreasing growth rate. The corresponding size $S$ of the growing entity approaches asymptotically a maximum value of

$$S_\infty = \frac{a}{|b|}.$$   (32)

The eqn (32) defines the mathematical *limit to growth*, which is often described incorrectly as the *carrying capacity* but it is only the carrying *capacity* if parameters $a$ and $b$ are clearly and convincingly related to the well-defined and well-explored ecological limits; otherwise, the calculated limit $S_\infty$ is just the calculated limit to growth, which may or may not represent the carrying capacity.

For instance, if we consider the growth of the GDP and if we determine empirically the parameters $a$ and $b$ using the *empirical values* of the growth rate it would be incorrect to claim that the calculated $S_\infty$ represents the empirically determined *carrying capacity* because the past economic growth might be following an unsafe trajectory and the economic collapse might happen even before reaching the calculated limit $S_\infty$. For this reason, describing the logistic limit as the carrying capacity may be misleading and it would be perhaps better to avoid such descriptions.

The same comment applies also to the calculated maximum when using the eqn (10). The calculated maximum, even if based on using the empirically-determined parameters $a$ and $b$, is just the calculated maximum. It also does not describe the carrying *capacity*. With limited resources the growth might be terminated even before reaching the maximum calculated using the empirically-determined parameters.

If $b > 0$ then, according to the eqn (29), the growth approaches singularity (escapes to infinity) at the time

$$t = t_s = -\frac{1}{a} \ln \frac{b}{aC}. \tag{33}$$

This type of growth resembles hyperbolic growth, which characterises the historical economic growth and the growth of human population (von Foerster, Mora & Amiot, 1960; Nielsen, 2014, 2016c, 2016d, 2016e). Hyperbolic growth (or to be more precise, the first-order hyperbolic growth) is given by the following simple equation:

$$S = (C - bt)^{-1}, \tag{34}$$

where $b > 0$.

Hyperbolic growth escapes to infinity when

$$t = t_s = \frac{C}{b}. \tag{35}$$

**Table 1.** *Fundamental equations for using linear approximations for growth rates in forecasting of trends.*

| Liniar Approximation | The Corresponding Distribution | Comments |
|---|---|---|
| $\frac{1}{S}\frac{dS}{dt} = a + bt$ | $S = C\exp\left[at + 0.5bt^2\right]$ | $S$ reaches a maximum when $a + bt = 0$ |
| $\frac{1}{S}\frac{dS}{dt} = bS$ | $S = (C - bt)^{-1}$ | If $b > 0$, hyperbolic growth. Singularity at $t_s = C/b$. Reciprocal values, $1/S$, decrease linearly with time. |
| $\frac{1}{S}\frac{dS}{dt} = a + bS$ | $S = \left[Ce^{-at} - \frac{b}{a}\right]^{-1}$ | If $b > 0$: pseudo-hyperbolic growth. Singularity at $t_s = -\frac{1}{a}\ln\frac{b}{aC}$. Reciprocal values decrease non-linearly to $-b/a$ when $t \to t_\infty$. If $b < 0$: logistic growth. $S$ increases asymptotically to $a/|b|$ when $t \to t_\infty$. |

Hyperbolic distribution is a solution of the following differential equation:

$$\frac{1}{S}\frac{dS}{dt} = bS. \tag{36}$$

If we compare this equation with the eqn (31) we can see that they are similar. In both cases, for $b > 0$, growth rate increases linearly with the size of the growing entity. However, while for the hyperbolic growth [eqn (36) with $b > 0$] the growth rate is *directly* proportional to $S$, for the growth described by the eqn (31) the linearly-increasing growth rate is displaced by the parameter $a$. It is a small difference but with significant consequences for the reciprocal values.

For the distribution described by the differential eqn (36) and by its solution (34), reciprocal values decrease linearly with time and they can be used as an uniquely identifying feature of hyperbolic distributions (Nielsen, 2014).

For the distribution described by the differential equation (31) and its solution (29), and for $b > 0$, reciprocal values of the size of the growing entity do not decrease linearly with time. They approach asymptotically the limit of $-b/a$.

However, graphically, solutions given by the eqns (29) and (34) look similar. They both increase slowly over a long time and fast over a short time and they both increase to infinity at a fixed time. Consequently, the solution given by the eqn (29) could be called a *pseudo-hyperbolic* distribution.

The curious difference between the respective differential equations, (31) and (36), is that the eqn (31) cannot be treated as the generalisation of the eqn (36). The two equations have to be solved independently. The solution to the eqn (31) cannot be used to derive the solution to the eqn (36). While solving the eqn (31) is difficult, solving the eqn (36) is simple. Its solution can be obtained by substitution $S = Z^{-1}$.

## 3. Mathematical method – substitutions

Fitting data and projecting growth can be also carried out by replacing $S$ or the growth rate $R$ by a suitably defined function and then checking whether such substitutions can be described by a linear approximation. A few examples are shown in Table 2.

The aim here is again to look for the simplest mathematical descriptions of growth rates. If the mathematical description of the growth rate of $S$ is complicated, it might be possible that the mathematical description of the growth rate of $F(S)$ could be simpler. Analysis of data can be simplified by looking for their alternative representations and the general idea is to try to reduce the analysis, if possible, to the simplest mathematical expression – the straight line.

Thus, for instance, if $F \equiv \ln S$, where $S$ represents the empirically-determined size of the growing entity, and if

$$\frac{1}{F}\frac{dF}{dt} = a + bt \,, \tag{37}$$

then

$$F = C\exp(at + 0.5bt^2)\,, \tag{38}$$

and

$$S = C\exp\left[\exp(at + 0.5bt^2)\right]. \tag{39}$$

The new constant C is now different than in the eqn (38) but it does not matter. It is a normalisation constant, which is determined by comparing calculated distribution with data.

If $F \equiv \ln S$ and if

$$\frac{1}{F}\frac{dF}{dt} = a + bF \,, \tag{40}$$

then

$$F = \left(Ce^{-at} - \frac{b}{a}\right)^{-1}, \tag{41}$$

and

$$S = \exp\left[\left(Ce^{-at} - \frac{b}{a}\right)^{-1}\right]. \tag{42}$$

Mathematical representations of $S$ given by the eqns (39) and (42) are not simple but they are acceptable because they are based on reducing mathematical analysis of data to the simplest representation given by a straight line for the growth rate of $F$.

We can also extend these alternative representations by replacing *the growth rate R* by a suitably defined function $F(R)$. If the mathematical description of the growth rate of $S$ turns out to be complicated it might be possible that a suitably-defined function $F(R)$ could simplify the analysis.

Thus, for instance, visual examination of the empirical growth rate of $S$ might suggest that it depends hyperbolically on time. We might try to fit hyperbolic distribution to the empirically-determined growth rate but it is also a good idea to check whether the distribution is indeed hyperbolic by examining the reciprocal values of $R$ because if $1/R$ should follow a straight line, then the distribution is hyperbolic (Nielsen, 2014). So, if

$$F(R) = \frac{1}{R} = a + bt ,$$ (43)

then

$$R = \frac{1}{S}\frac{dS}{dt} = \frac{1}{a+bt} .$$ (44)

Hyperbolic distribution is not as simple as a straight line but it can be reduced to a straight line, which is easier to accept and understand. Such an exercise increases confidence that the distribution is indeed hyperbolic or at least that it can be well approximated by a hyperbolic distribution.

The differential equation (44) can be presented as

$$\frac{dS}{S} = \frac{dt}{a+bt} ,$$ (45)

which, when integrated, gives

$$\ln S = \frac{1}{b}\ln(a+bt) + C .$$ (46)

Consequently,

$$S = C(a+bt)^{1/b}$$ (47)

because $\exp(\ln z) = z$. The constants $C$ are different in these last two equations but again it does not matter because they are just the normalisation constants, which have to be determined by comparing calculated $S$ with its corresponding empirical value.

The eqn (47) is not simple but it has been obtained by reducing mathematical analysis to the simplest mathematical expression given by the eqn (43), which identifies hyperbolic distribution of $R$. The fundamental starting step is simple and the derived expression for $S$, even if complicated, can be accepted with a high degree of confidence.

If a visual examination of the empirical growth rate $R$ suggests that it follows an exponential distribution, we can try to fit an exponential function to $R$ or to display it using the semilogarithmic scales of reference. If

$$\ln R = a + bt ,$$ (48)

then

$$\frac{1}{S}\frac{dS}{dt} = \exp(a + bt) \qquad (49)$$

and the solution to this equation is

$$S = C\exp\left[\frac{e^a}{b}e^{bt}\right] \qquad (50)$$

Again, it is not a simple description of $S$ but this complicated expression has been derived using the simplest representation of $R$ via $\ln R$.

All mathematical descriptions of $S$ presented here [eqns (10), (29), (34), (39), (42), (47) and (50)] are not simple but all of them were derived using the simplest mathematical representations of related quantities. It is easy to *construct* complicated and dubious formulae but even complicated formulae are acceptable if they are *derived* using simple and acceptable assumptions.

If the growth rate of $S$ is represented directly by an exponential distribution, i.e. if

$$\frac{1}{S}\frac{dS}{dt} = ae^{bt} \quad , \qquad (51)$$

then, using the general eqn (6), we can find that

$$S = C\exp\left(\frac{a}{b}e^{bt}\right). \qquad (52)$$

Solutions given by eqns (50) and (52) are the same. The difference is only in the way parameters $a$ and $b$ are defined. The advantage of using the linear representation given by the eqn (48) is that it allows for a clear demonstration whether growth rates follow exponential distribution or not. Exponential description of data applies, for instance, to the growth rate between 1963 and 2017 describing the growth of the world population.

**Table 2.** *Examples of extended applications of linear approximations to describe complicated distributions.*

| Linear Approximation | The Corresponding Distribution |
|---|---|
| $F \equiv \ln S$ ; $\dfrac{1}{F}\dfrac{dF}{dt} = a + bt$ | $S = C\exp\left[\exp(at + 0.5bt^2)\right]$ |
| $F \equiv \ln S$ ; $\dfrac{1}{F}\dfrac{dF}{dt} = a + bF$ | $S = \exp\left[\left(Ce^{-at} - \dfrac{b}{a}\right)^{-1}\right]$ |
| $F \equiv \dfrac{1}{R} = a + bt$ ; $R \equiv \dfrac{1}{S}\dfrac{dS}{dt}$ | $S = C(a + bt)^{1/b}$ |
| $\ln R = a + bt$ ; $R \equiv \dfrac{1}{S}\dfrac{dS}{dt}$ | $S = \exp\left[\dfrac{e^a}{b}e^{bt}\right]$ |

We could also have other examples of growth rates, whose description could be reduced to linear approximations. For instance, analysis of the world Gross Domestic Product indicated that the growth rate followed a familiar mathematical distribution (Nielsen, 2015a).

$$R \equiv \frac{1}{S}\frac{dS}{dt} = (a - be^{-rt})^{-1} \qquad (53)$$

This distribution can be also reduced to a linear distribution by using the following equation

$$F \equiv \ln\left(a - \frac{1}{R}\right) = \ln b - rt . \qquad (54)$$

This might sound like making it more complicated but it is not because, as mentioned before, deviations of data from the straight line can be easily detected and using straight lines could be used as a convenient test whether our mathematical interpretation of growth rate data is correct. In the case of global economic growth, the assumption that the growth rate should be described by the eqn (53) is correct has been confirmed by the linear distribution given by the eqn (54).

The solution of the differential eqn (53) is

$$S(t) = C\exp\left[\frac{t}{a} + \frac{1}{ra}\ln\left(a - be^{-rt}\right)\right]. \qquad (55)$$

This distributions changes asymptotically into exponential distribution with the growth rate of $1/a$.

## 4. Examples

It is important to understand that only general trends of growth rates determine the general trends of the corresponding growth trajectories. Thus, only general trends of growth rates should be used in projecting growth. Fine structures, such as moderate oscillations or fluctuations can be ignored because they have no impact on the general trends of growth.

Furthermore, in reproducing the general trends of growth rates it is essential to use functions, which do not depend critically on the range of data. Suitable functions are straight line, exponential function or any other function that can be reduced to a straight line, such as the function described by the eqn (53). Higher order polynomials should not be used. They are suitable for *reproducing* growth trajectories or for using them in polynomial interpolations of gradients but they are not suitable in forecasting because their shape depends strongly on the range of data.

Application of differential equations in the analysis of trends and in forecasting is explained in Figures 1-3.

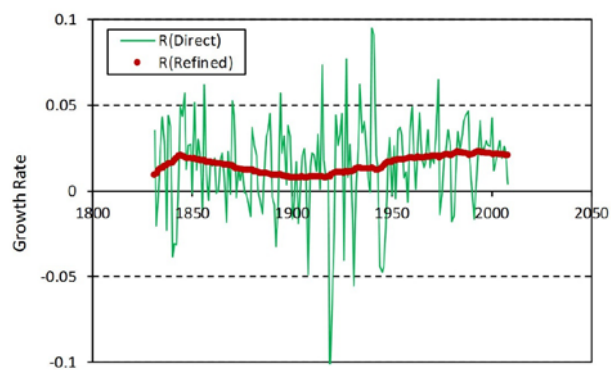

**Figure 1.** *Two sets of calculations of the growth rate of the Gross Domestic Product per capita (GDP/cap) in the United Kingdom, based on using data of Maddison (2010), are presented. R(Direct) is the growth rate calculated directly from the GDP/cap data using the eqn (2). R(Refined) is calculated using data for the GDP/cap and a smoothed-out gradient. It shows the fine structure, which was obscured by fluctuations of R(Direct).*

Figure 1 shows the growth rate of the Gross Domestic Product per capita (GDP/cap) in the United Kingdom between 1830 and 2008 calculated using Maddison's data (Maddison, 2010). There are two sets of calculations in this figure, identified as R(Direct) and R(Refined).

R(Direct) is the growth rate calculated directly from data using the eqn (2). Results of such calculations are influenced strongly by local gradients. Small deviations in the empirical values of the GDP/cap can produce large differences in the calculated growth rate. Conversely, large fluctuations in the growth rate can be expected to be reflected in only small variations in the data describing growth trajectory.

R(Refined) is the growth rate calculated using gradients smoothed out by polynomial interpolation. The noise created by random local gradients is then filtered out and a clear trend is revealed. We can see now that the growth rate was in fact gently oscillating. These gentle oscillations were obscured by strong fluctuations of R(Direct).

Our first step now is to use the discrete values of the growth rate, calculated from data, either directly or by using interpolated gradient, and to calculate the growth trajectory of the growing entity, which in our case is the GDP/cap. To this end, we have to carry out numerical integration of the following differential equation:

$$\frac{1}{S}\frac{dS}{dt} = R_e \,, \tag{56}$$

where $R_e$ is represented by the discrete values of either R(Direct) or R(Refined).

Results of these calculations are shown in Figure 2. The top section of these figure presents the full range of data compared with the results of the numerical integration of the eqn (56). Unfortunately, results of calculation are obscured by data. To see them better we have to display a smaller section of data as shown in the lower part of Figure 2. We can see now that small ripples in the growth trajectory of the GDP/cap are associated with large fluctuations in the growth rate R(Direct) calculated directly from data. The general trend of the GDP/cap is well reproduced by the gently varying growth rate, R(Refined).

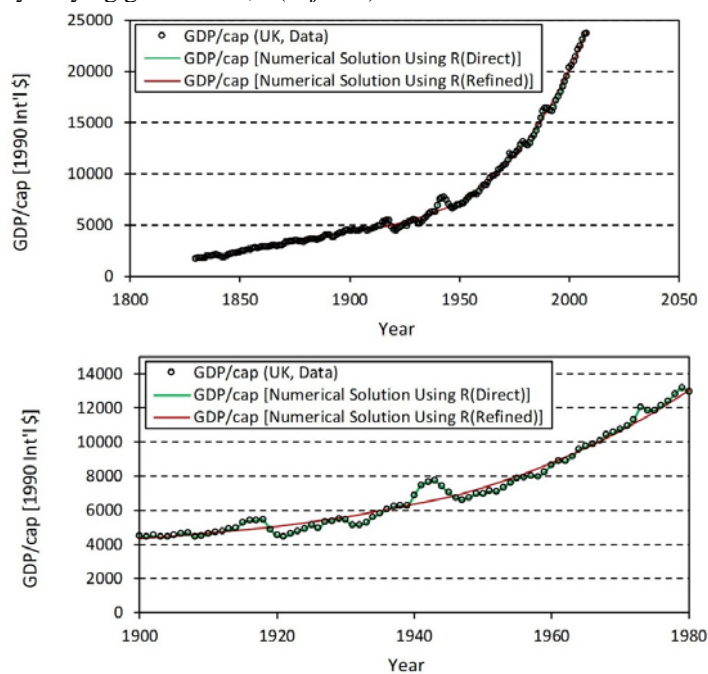

**Figure 2.** *Results of numerical integration of the eqn (56) using growth rates R(Direct) and R(Refined) shown in Figure 1. The Gross Domestic Product per capita (GDP/cap) is expressed in the 1990 International Geary-Khamis dollars. Results of calculations are obscured by data. To see them more clearly, a smaller section of the range of data is shown in the lower part of this figure. The green line represents the calculated curve using the fluctuating values of R(Direct). It is not a line drawn through the data.*

Our next step is to see how the analytical representations of the growth rate are reflected in the growth trajectory of the GDP/cap. To this end, we have to solve analytically the following differential equation:

$$\frac{1}{S}\frac{dS}{dt} = f(t) , \qquad (57)$$

where, in our case, $f(t)$ is either a straight line fitted to $R(Refined)$, as shown in the top part of Figure 3, or the sixth order polynomial reproducing gentle oscillations of $R(Refined)$. The best linear fit to $R(Direct)$ is virtually the same as the best linear fit to $R(Refined)$. These distributions are given by the following equations:

$$f(t) = a + bt \qquad (58)$$

and

$$f(t) = \sum_{i=0}^{6} a_i t^i . \qquad (59)$$

Their parameters are: $a = -8.964 \times 10^{-2}$ , $b = 5.459 \times 10^{-5}$ , $a_0 = -1.412 \times 10^6$ , $a_1 = 4.338 \times 10^3$ , $a_2 = -5.551 \times 10^0$ , $a_3 = 3.787 \times 10^{-3}$ , $a_4 = -1.453 \times 10^{-6}$ , $a_5 = 2.974 \times 10^{-10}$ and $a_6 = -2.535 \times 10^{-14}$ .

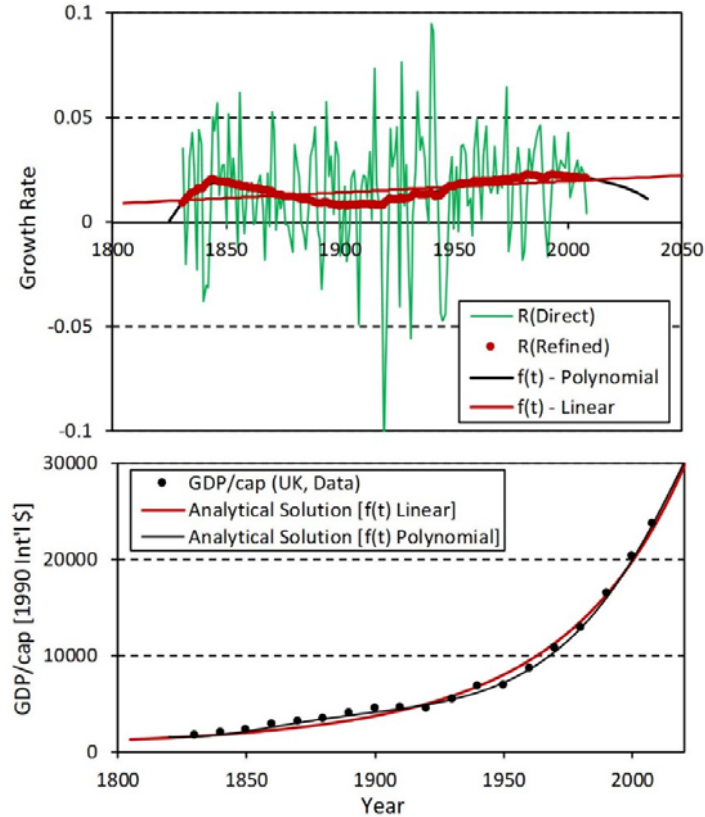

**Figure 3.** *Analytical solutions of the eqn (57) are shown in the lower part of this figure, with $f(t)$ represented either by a straight line or by the sixth-order polynomial fitted to R(Refined), as shown in the top part. The straight-line fit to R(Direct) is virtually the same as the straight-line fit to R(Refined). To display the difference between these two analytical solutions, data in the lower section are shown in steps of 10 years. While the solution corresponding to the sixth-order polynomial reproduces gentle oscillations in the growth trajectory of the GDP/cap, the solution corresponding to the straight-line representation of the growth rate reproduces the general trend of the economic growth and thus is perfectly suitable for predicting the growth trajectory.*

Solutions of the eqn (57) are given by the eqn (6). They are shown in the lower part of Figure 3. The solution corresponding to the sixth-order polynomial follows the data precisely and reproduces the small oscillations in the growth trajectory. Unfortunately, this solution cannot be used in forecasting because the calculated trajectory depends strongly on the range of data, as we can see in the top part of Figure 3. The growth rate described by the sixth-order polynomial is only realistic strictly within the range of the growth rate data.

So, we are left with only one solution, which does not depend critically on the range of growth rate data, the solution corresponding to the straight line describing the growth rate (see the top part of Figure 3). As we can see in the lower part of Figure 3, this solution gives excellent description of the growth trajectory and can be used in forecasting.

In general, as long as $R(Refined)$ oscillates gently around, or follows closely, a straight line, or some other function, which does not depend critically on the range of data, such simplified representations of $R(Refined)$ can be used successfully not only in the description of the time-dependent series but also in the reliable forecasting. However, for large oscillation, projections of trends are unreliable. A trend has to display a certain degree of stability to be predictable.

Now that we understand which features of the growth rate are important in forecasting, we can use an example of the growth of the world population. The advantage of using this example is that we can compare our calculation with the independent calculations carried out recently by the United Nations (2015).

The top part in Figure 4 shows the growth rate for the growth of the world population calculated directly from the population data (US Census Bureau, 2017). In this case, the data were of such good quality that the interpolation of gradients was unnecessary.

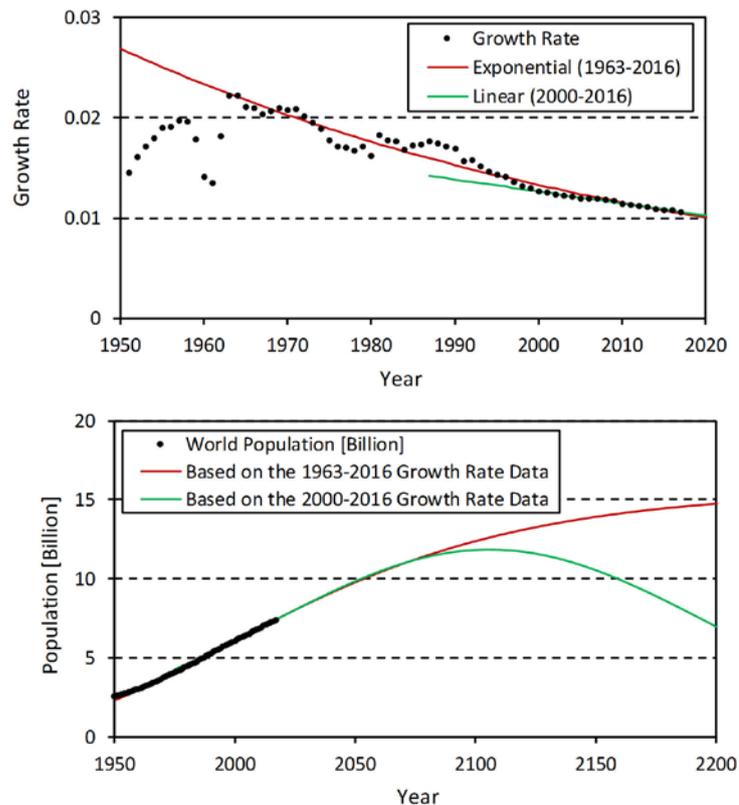

**Figure 4.** *Forecasting of the growth of the world population. Two representations (exponential and linear) of the growth rate based on the data of the US Census Bureau (2017), are used to generate growth trajectories for the growth of the world population. These calculations are in good agreement with projections of the United Nation (2015). The UN publication gives no information about the growth of population in the 22nd century.*

In order to project growth, we seem to have two obvious options: (1) to use the wide range of growth rate data between 1963 and 2016, which can be well described by exponential function or (2) to assume that from around 2000 the growth rate is now settling along a linearly decreasing trajectory. The projection of growth of the world population based on fitting exponential distribution to the growth rate could be considered as more reliable because it is based on a wide range of data but it is still possible that the growth rate will now follow a linearly decreasing trajectory.

Results of calculations are shown in the lower part of Figure 4. The trajectory corresponding to the exponentially decreasing growth rate is given by the eqn (52). It is a pseudo-logistic trajectory because it increases asymptotically to a constant value given by the normalization constant $C$. Its parameters are: $a = 2.179 \times 10^{10}$ and $b = -1.406 \times 10^{-2}$ and its asymptotic value is 15.6 billion.

The trajectory corresponding to the linear fit to the growth rate is given by the eqn (10). Its parameters are: $a = 2.520 \times 10^{-1}$ and $b = -1.197 \times 10^{-4}$. It reaches a maximum of 12.4 billion in 2106.

Calculations shown in Figure 4 are in good agreement with projections of the United Nations (2015). According to this source "The world population is projected to increase by more than one billion people within the next 15 years, reaching 8.5 billion in 2030, and to increase further to 9.7 billion in 2050 and 11.2 billion by 2100" (United Nations, 2015, p. 2). My prognosis is 8.4 billion in 2030, 9.8 billion in 2050 and 11.8 billion in 2100 for the trajectory leading to the localized maximum. If the growth of the world population is going to follow the trajectory leading to the asymptotic maximum, then it will also reach 8.4 billion in 2030 and 9.8 billion in 2050 but only a slightly larger size of 12.4 billion in 2100. The difference between predicted values in 2100 is so small that we shall not know until the next century whether we are likely to reach a localized maximum of around 12 billion or to have the population continually increasing to the asymptotic size of around 15.6 billion, if such a large size can be supported by the accessible resources.

A summary of all these predictions is presented in Table 1. The UN prediction gives no information about the growth of population in the 22nd century. For the 21st century, the agreement between these two independent calculations is remarkably close.

**Table 1.** *Predicted growth of the world population*

| Source | 2030 | 2050 | 2100 | $S_{max}$ | $S_a$ |
|--------|------|------|------|-----------|-------|
| UN | 8.5 | 9.7 | 11.2 | NI | NI |
| CA | 8.4 | 9.8 | 11.8 | 11.9 | NA |
| CA | 8.4 | 9.8 | 12.4 | NA | 15.6 |

UN – United Nations, 2015; CA – current analysis; NI – no information; NA – not applicable; $S_{max}$ – maximum value; $S_a$ – asymptotic value

These calculations should be taken as the prediction of the most likely future, which can be still changed. The future depends on our actions and a safer future would be in a lower size of the projected population. It is in our power to do it but we would have to work harder on reducing the growth of global population. The obvious place of our attention should be in poorer countries, and the way to do it is to improve their standard of living (Nielsen, 2016f).

Another good example of the application of the mathematical method of forecasting described in the presented here document is the economic growth in Japan because, as we shall soon see, the future is already here and thus we can use data to see how reliable is our forecasting.

Figure 5 presents the growth rate of the GDP in Japan as the function of time. We can see that the growth rate, as described by $R$(*Refined*) was decreasing but then, around 2007, it started to increase. However, between 1975 and 2007, it followed an approximately linearly decreasing trajectory. We can ask, therefore, what would have happened if the growth rate in Japan continued to follow this linearly decreasing trajectory and the answer is that it would have reached a maximum because the straight line crosses the horizontal axis around 2007. From around that time, the GDP in Japan would start to decrease. However, the growth rate came close to zero and started to increase indicating that rather than reaching a maximum, the growth of the GDP in Japan continued to increase, which is hardly surprising because whenever possible, negative growth rate is always avoided.

We can also check whether the growth rate of the GDP followed a logistic trajectory by plotting it as the function of the size of the GDP. Such a plot is presented in Figure 6, which shows that the growth rate was indeed decreasing linearly over a long time and thus that the growth of the GDP might have been logistic. We should understand that the distinction between alternative trajectories is not immediately obvious. Growth trajectories corresponding to different representations of the growth rate are indistinguishable over a long time. Only when the growth reaches a maximum or approaches it asymptotic value we might see distinction between two alternative trajectories. We can see this feature in the lower part of Figure 4 and we shall soon see the same feature for the growth of the GDP.

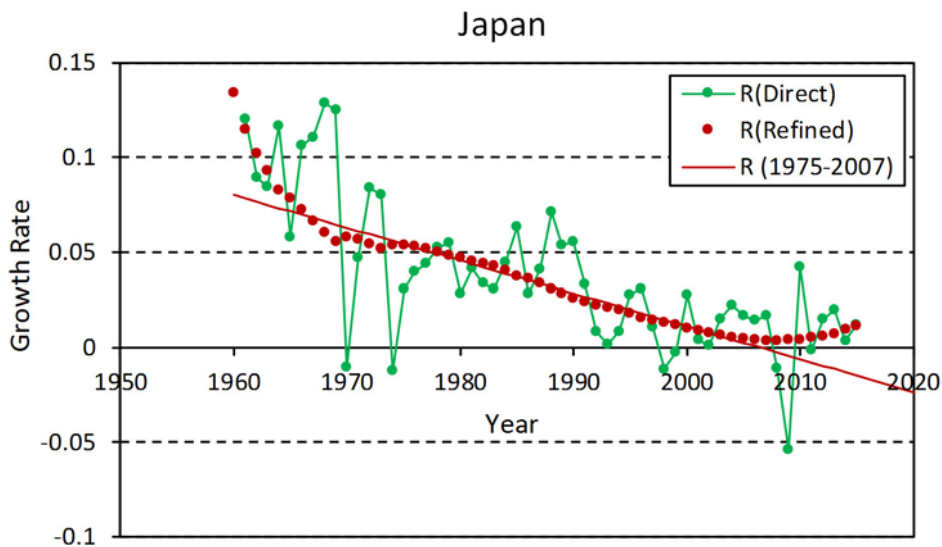

**Figure 5.** *Growth rate of the GDP in Japan as a function of time. The data are from the World Bank (2017).*

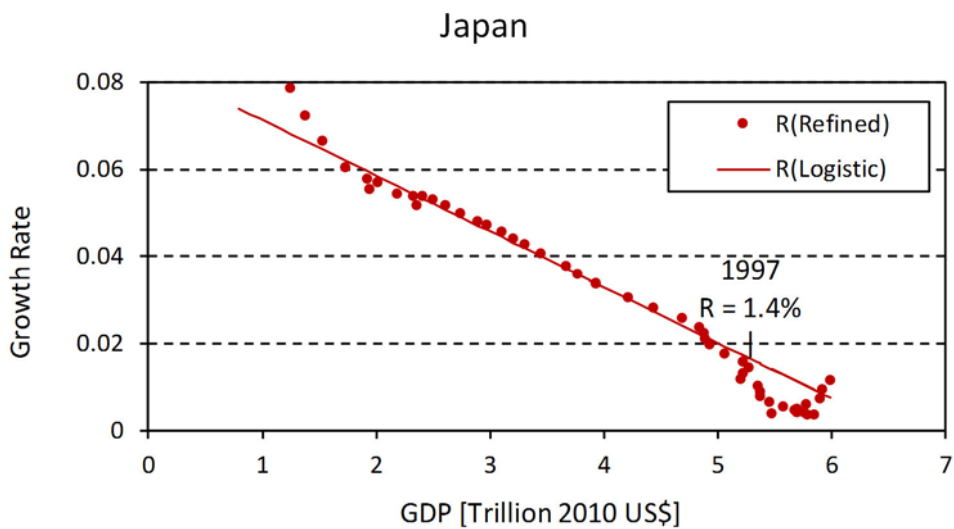

**Figure 6.** *Growth rate of the GDP in Japan as a function of the size of the GDP. The data are from the World Bank (2017). When the growth rate decreased to 1.4% it became unstable. The same process occurred in Greece (Nielsen, 2015b). Japan should not make the same mistake as it was made in Greece by trying to boost its growth rate. For the safe economic growth, the growth rate of the GDP in Japan should be now maintained between zero and 1%.*

It is important to notice that when the growth rate reaches a low value, economic growth is likely to be unstable. If it approaches a maximum, then it should soon change to a negative value, but no country in the world would be happy to have a consistently negative growth. Consequently, it can be expected that all efforts will be made to keep the growth rate positive. If the growth of the GDP approaches its asymptotic value, then it should decrease asymptotically to zero. Such a fine tuning is practically impossible and the growth rate might vary randomly around a small positive value or it might be forced to increase substantially and thus to depart from its logistic trajectory. Such a situation happened in Greece (Nielsen, 2015b). Their growth rate was decreasing fast along a size-dependent linear trajectory, reached a low value, became unstable, was forced to increase along a linear trajectory and inevitably led to the economic collapse.

We can now use the straight-line trajectories presented in Figures 5 and 6 to project the economic growth in Japan. Results are shown in Figure 7.

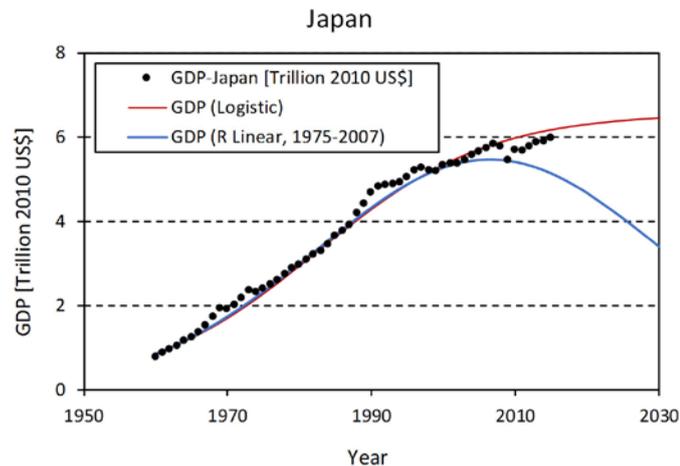

**Figure 7.** *Growth of the GDP in Japan. Projected growth is compared with data. The GDP values are close to the predicted logistic trajectory. Economic growth in Japan will remain sustainable as long the growth rate is going to be kept above zero but below around 1%.*

We can see that, as mentioned earlier, for a long time there is no distinction between the logistic trajectory and the trajectory leading to a localized maxim. It is only when the growth is close to the localized maximum that the distinction between the two trajectories becomes clear.

The logistic trajectory is described by the eqn (29) with parameters $a = 8.411 \times 10^{-2}$ and $b = -1.279 \times 10^{-2}$. Its asymptotic value is $6.573 \times 10^{12}$ (2010 US$). The trajectory leading to a localized maximum is described by the eqn (10) with parameters $a = 3.452 \times 10^{0}$ and $b = -1.726 \times 10^{-3}$. Its maximum was $5.469 \times 10^{12}$ in 2006. Figure 7 shows that the forecasting of growth based on the described method gives again reliable results. Not only does it describe the growth when alternative trajectories are indistinguishable but also projects reliably the growth when the distinction is well pronounced. Economic growth in Japan follows now closely the logistic trajectory. As mentioned earlier, it is difficult to follow precisely this trajectory because fine tuning of the growth rate is required. However, to follow it precisely is not critical. What is critical, as indicated by these calculations, is to keep the growth rate at a low level. A small constant value would give a slow exponential growth, which might be, for a long time, close enough to the logistic trajectory. On no account, the growth rate should be forced to increase above 1%, unless only temporarily.

## 5. Summary

I have described mathematical method of analysis of growth rates and of predicting growth trajectories. Growth rate can be presented either as a function of time or as a function of the size of the growing entity. The general aim is to find a linear representation of the growth rate and then use differential equations to translate the linearly represented growth rate into mathematically more complicated descriptions of growth trajectories, which in turn can be used in forecasting. Other representation, which do not depend critically on the range of data can be also used. The simplest of these alternative representation is the exponential function. Another example is a function described by the eqn (53). Polynomial descriptions of growth rates should not be used because they depend strongly on the range of data. They are unsuitable in forecasting of trends but they are suitable for more refined representations of growth rates. The method described in this document is illustrated by numerous examples listed in Tables I and II and in Figures 1-7. It is a simple method, which is easy to use.

Forecasting of trends can be used as the essential tool in shaping the future. It can show what steps should be taken to control growth and how to avoid undesirable outcomes. Not all critical trends can be successfully altered but if we can understand their dynamics we might be better equipped to search for successful solutions.